\documentclass[doublecol]{epl2} 
\usepackage{epsfig}

\title{Intricate dynamics of a deterministic walk confined in a strip}
\shorttitle{Intricate dynamics of a deterministic walk confined in a strip}  

\author{Denis Boyer\inst{1}}
\shortauthor{D. Boyer}

\institute{                    
  \inst{1} Instituto de F\'\i sica, Universidad
  Nacional Aut\'onoma de M\'exico, Apartado Postal 20-364, 
01000 M\'exico D.F., M\'exico
}

\abstract{We study the dynamics of a deterministic walk confined
in a narrow two-dimensional space randomly filled with point-like targets. 
At each step, the walker visits the nearest target not previously visited. 
Complex dynamics is observed at some intermediate values of the domain 
width, when, while drifting, the walk performs long intermittent 
backward excursions. 
As the width is increased, evidence of a transition from 
ballistic motion to a weakly non-ergodic regime is shown, characterized 
by sudden inversions of the drift velocity with a probability slowly 
decaying with time, as $1/t$ at leading order. Excursion durations, 
first-passage times 
and the dynamics of unvisited targets follow power-law distributions.
For parameter values below this scaling regime, precursory patterns in 
the form of \lq\lq wild" outliers are observed, in close relation with the 
presence of log-oscillations in the probability distributions.
We discuss the connections between this model and several evolving 
biological systems.}

\begin{document}

\maketitle

\section{Introduction}

Deterministic walks in disordered environments have received an 
increasing attention over the past years. They describe diffusion 
processes following non-random rules and have applications, among 
others, to the study of the displacements 
of individuals in complex landscapes. Examples are human travels 
\cite{lima,stanley}, human displacements in a city \cite{chowell}, 
movement patterns 
of hunter-gatherer \cite{juhoansi} or foraging animals \cite{boyer}.  

From a given position, the next site visited by a purely deterministic walker 
is assigned from a given set of rules and not stochastically. These walks still 
have probabilistic and fluctuating features if the environment is random or 
heterogeneous. Interesting dynamics have been observed, such 
as normal \cite{boon} or anomalous diffusion \cite{redqueen}, behaviors 
analogous to that of the Lorentz gas \cite{bunimovich}, cycles with power-law 
distributed periods \cite{lima,derrida} or L\'evy-like step length 
distributions \cite{physicaA,santos}. Complex behavior can emerge from very 
simple rules, {\it e.g}, when each individual step 
optimizes a given cost function. Some properties of deterministic walks 
have also been used as tools to process large data sets in
galaxy surveys \cite{elson}, thesaurus graphs \cite{kinouchi} 
or for pattern recognition \cite{campiteli}.

In many situations, in particular biological, the deterministic walker itself 
changes the medium, which introduces memory \cite{bunimovich}. 
An important case 
is the self-avoiding walk (SAW), which can be implemented to model biological 
systems with negative feedbacks that tend to avoid past behaviors. A simple
example is that of a foraging animal relying on mental maps to navigate an 
environment composed of food patches that are not revisited after they have
been depleted \cite{boyer}. In a different context, the brain activity has 
been modeled by random walks keeping memory of their complete history
in order to avoid persistent patterns; recent memory loss producing 
pathological repetitions, like in the Alzheimer's disease \cite{gandhi}. In 
evolutionary ecology, the well-known Red-Queen principle assumes that any 
organism must constantly evolve in order to prevent 
its predators or preys to adapt to an otherwise predictable behavior. 
Similar considerations can apply to the dynamics of technological 
innovations \cite{kauffman}.
Freund and Grassberger introduced some time ago a self-avoiding deterministic 
walk model in disordered two-dimensional domains, mimicking evolving organisms 
in phenotype landscapes \cite{redqueen}. 
These kinds of models are very difficult to handle analytically; 
they are firstly dynamical and usually not equivalent to canonical 
SAWs \cite{indous}.

Here, we study a minimal model of a deterministic walk with a SAW constraint 
(in the infinite memory limit) and confined in a nearly one dimensional random 
medium. In an evolutionary context, whereas SAWs can be justified 
by natural selection, organisms also have developmental
constraints due to limited phenotypic variability \cite{arnold}.
Similarly, the development of human artifacts is restricted by design 
limits. We model this important constraint by a narrow random 
medium where the walker can evolve without bounds only in one 
direction. In narrow landscapes, the model exhibits very rich dynamical 
features not observed in unbounded ones, such as intermittent behavior,
scaling laws, discrete scale invariance and very large events (outliers). 
A discussion of these results is then presented.

\section{Model description}
Consider a two-dimensional strip of width $l$ and infinite 
length along the horizontal direction. The strip is randomly 
filled with fixed point-like targets with uniform number density $\rho_0$,
representing, say, food patches for a foraging animal or
phenotypes for an evolving species. The only control parameter
is the reduced domain width, defined as $\delta=l/l_0$, with 
$l_0=\rho_0^{-1/2}$
the characteristic distance between neighboring targets. 
At time $t=0$, a walker is located at some target 
with coordinates ($x_0,y_0$), taken as the origin. 
Two rules of motion are then recursively 
applied: the walker $(i)$ moves to the nearest available target, $(ii)$ 
does not visit a previously visited target. When the new target is reached, 
$t$ is updated to $t+1$.

The medium can be made one(two)-dimensional in the limit $\delta\ll 1$ 
($\delta\gg 1$),
respectively. We will focus here on values of $\delta$ of $O(1)$, 
typically in the range $(2,5)$, such that
the walker has a some vertical degree of freedom but
a practically one-dimensional motion on large scales,                        
described by its horizontal coordinate $x(t)$.
In the simulations,
the medium is a rectangle of area unity containing $N$ targets and of width
$l=\delta/\sqrt{N}$. Each run start near the middle of the domain and 
is stopped before the walker reaches the lateral vertical walls.

\section{Trajectories}
In the one dimensional case ($\delta\ll 1$), the targets are randomly 
distributed on a line and the motion is simply ballistic. 
After a possible short transient, the walker breaks the right-left symmetry
and always moves to the nearest target to its right (or left) so 
that $x(t)$ is a sum of same-sign independent random variables with Poisson
distribution. 
The $2d$ case ($\delta\gg1$) is sometime called the 
\lq\lq tourist walk" \cite{lima,stanley}: the
trajectories are not very different from $2d$ random walks, although
slightly superdiffusive \cite{lopezcorona}.

\begin{figure}
\onefigure[width=7cm,angle=-90] {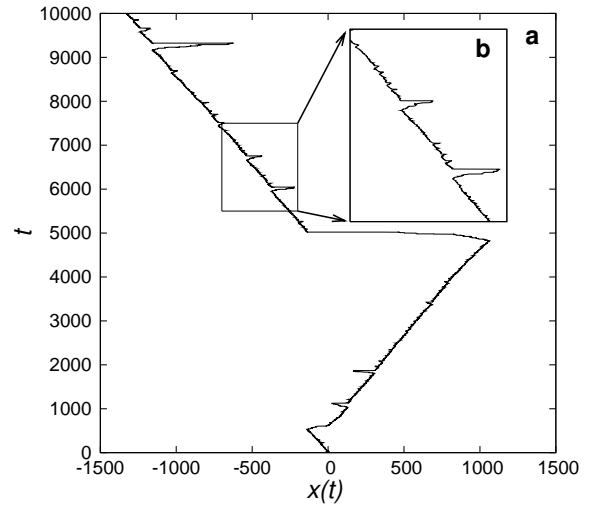}
\caption{Space-time diagram of a trajectory with $\delta=4.1$ ($x$ is 
in unit of $l_0$). }
\label{figTrajectory}
\end{figure}

For the cases $\delta=O(1)$ of interest here, the situation is quite different 
and trajectories exhibit a rich structure. 
As shown in fig.~\ref{figTrajectory} at $\delta=4.1$,
the motion is on average ballistic due to the 
confining effect of the horizontal walls. 
The numerically calculated root-mean-square displacement 
$\langle x(t)^2\rangle^{1/2}$ follows a linear behavior with time
(not shown here).
Note that the walker horizontal 
velocity $x(t)-x(t-1)$ often changes sign: the walker performs many
\lq\lq backward excursions" while drifting along the strip. 
These excursions, that were observed in a preliminary study of the 
model \cite{physicaA}, can be explained qualitatively. A walker drifting, 
say, toward 
the left does not necessarily visit all the targets of a given neighborhood 
on its way and may ignore some targets. From time to time, rules $(i)$ and 
$(ii)$ make the walker turn back and visit these unvisited targets toward 
the right, until it ends up in a region depleted of available targets. 
In that case, a single step can brings the walker back to the unexplored region 
located to the left. 

Unexpectedly,
backward excursions of all sizes can be observed in 
fig.~\ref{figTrajectory}a. Whereas most excursions are short, some can 
be of order $10^2-10^3\times l_0$, {\it e.g.} 
near $t=9300$. A close up of fig.~\ref{figTrajectory}a 
(inset b) reveals further details and suggests that the 
trajectory is fractal. In \cite{physicaA,santos}, it was found that these
intermittent backward excursions can lead to \lq\lq L\'evy-like" 
distributions for the distance separating successively visited targets,
of the form $\ell^{-(1+\mu)}$, with $\mu\simeq1$ at $\delta= 4$ \cite{santos}.

Additionally, the sign of the drift velocity can change suddenly at large $t$ 
({\it e.g.} at $t\approx600$ and $t\approx5000$
in fig.~\ref{figTrajectory}a). Such inversions
happen during a backward excursion, at some point when the closest unvisited 
target is located, say, to the right of $x_0$ for a trajectory that was 
previously drifting toward the left. Obviously, inversions can not
occur in the $1d$ ballistic limit of the model. We investigate below the 
possible existence of a transition between different dynamical regimes 
as $\delta$ is varied.

\section{Inversion probability and first-passage times}
We define the explored
interval at time $t$ as $[x_{min}(t),x_{max}(t)]$, where 
$x_{min}(t)$ ($x_{max}(t)$) is the coordinate of the leftmost (rightmost)
visited target after $t$ steps, respectively. A inversion 
(say, from right to left) occurs during the $t^{\rm th}$ step 
if $x_{min}(t)-x_{min}(t-1)<0$
and if there exists a time $t'<t$ such that $x_{max}(t')-x_{max}(t'-1)>0$ 
and such that $x_{min}(t")-x_{min}(t"-1)=x_{max}(t")-x_{max}(t"-1)=0$ 
for $t'<t"<t$.

We then define $P_{inv}(t)$ as the probability that an inversion (to the
left or right) occurs during the $t^{\rm th}$ step, and $P_0(t)$ 
as the probability that the walker crosses $x_0$ during the 
$t^{\rm th}$ step. In the random walk language, $P_0$ is analogous to the 
probability of presence at the origin. If motion is essentially ballistic 
between two inversions, a trajectory crosses 
its origin at large times only during an inversion: $P_{inv}$ 
and $P_0$ have the same asymptotic behavior.

\begin{figure}
\onefigure[width=7cm,angle=-90] {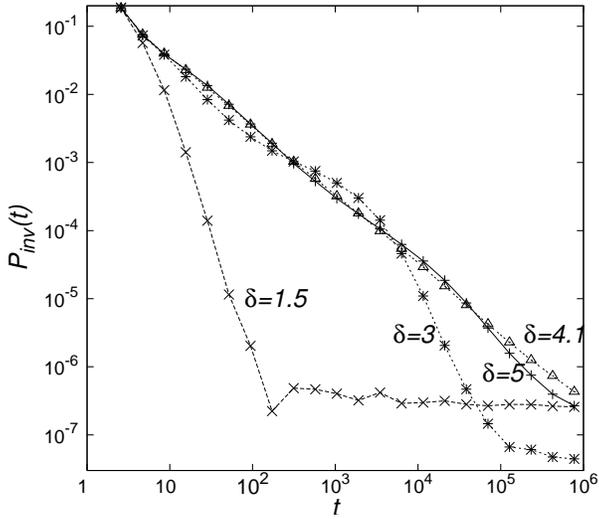}
\caption{Probability that the drift velocity changes its sign
at time $t$, as a function of $t$ and for various strip widths. The
probabilities are calculated from $8\ 10^4$ independent random media.}
\label{figPinv}
\end{figure}

\begin{figure}
\onefigure[width=7cm,angle=-90] {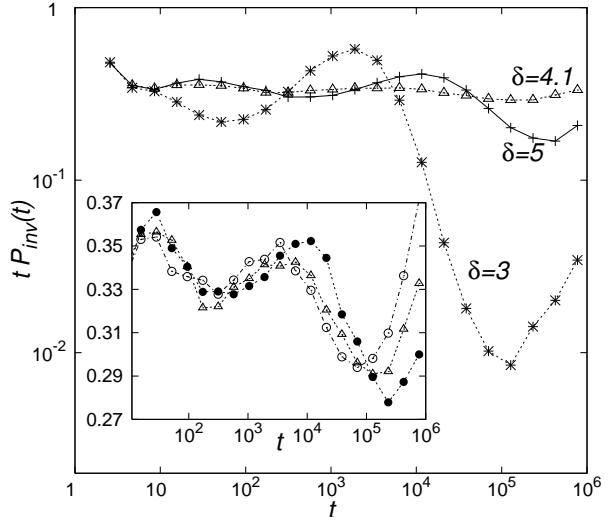}
\caption{Same data as in fig.~\ref{figPinv}, reploted as $tP_{inv}(t)$ vs. $t$.
Inset: details of $tP_{inv}(t)$ for $\delta=4$ ($\circ$), 4.1 ($\triangle$),
4.2 ($\bullet$).}
\label{figScalPinv}
\end{figure}

In Figure \ref{figPinv}, the numerically computed $P_{inv}(t)$
decays very slowly with time for several $O(1)$ values of the strip width 
$\delta$. In domains as narrow as $\delta=1.5$, after an initial
steep decay,  $P_{inv}(t)$ exhibits a surprising fat tail.
At the larger value $\delta=4.1$, $P_{inv}(t)$ 
can be well fitted by the simple inverse power-law $c/t$, with $c$ a 
constant. The same curves are reploted in Figure \ref{figScalPinv} as
$tP_{inv}(t)$ versus $t$: for $\delta=4.1$, the curve remains remarkably 
constant during almost 6 decades, while strong corrections to scaling
are present below and above that parameter value ($\delta=3$ and $5$). 
Very similar results are obtained for $P_0(t)$.

Despite that the walker crosses less frequently the origin than a $1d$
random walker (where $P_0(t)\sim t^{-1/2}$), the return probability 
at large times in narrow strips ($\delta\ll t$) remains very
high instead of being exponentially small as
for usual ballistic motion ({\it e.g.}, a $1d$ random walker with a bias).

Contrary to random walks, sign changes in $x(t)$ are abrupt and
not strongly correlated to the evolution of $x(t)$ during the preceding 
steps (see Fig. \ref{figTrajectory}). It is therefore useful to make a 
connection between this result
and a simpler two-state stochastic problem consisting of a 
walker moving ballistically on a line with two possible velocities, 
$v$ and $-v$. Starting in one state, the walker change its velocity in the 
time interval $[t,t+1]$ 
with probability $p_{inv}(t)$, that is given. It is well known that if 
$p_{inv}(t)$ decays faster than $1/t$, there is a finite probability 
that the walker remains indefinitely in a same state ($v$ or $-v$) after 
reaching this state. If $p_{inv}(t)$ decays as $1/t$ or slower, the 
probability that the walker remains in a same state forever is zero.
The behavior of the system is non-ergodic in the former case, as
the left-right symmetry is asymptotically broken, while it is weakly 
non-ergodic 
\footnote{In this context, weakly non-ergodic means that both states always 
remain accessible to the walker, although the time interval between two 
visits diverges asymptotically \cite{bouchaud}.} 
or ergodic in the latter case. Obviously, the above two-state problem only
provides an approximate description of our model and makes sense
only in the regime where inversions are abrupt. 

We investigate more in details the possibility of a 
non-ergodic/weakly non-ergodic
transition as the strip width $\delta$ is increased across some critical value 
$\delta_c$ where $P_{inv}(t)\simeq c/t$. Making an analogy between the 
behavior of $P_{inv}$ (or $P_0$) and that of a correlation function near a 
critical point, for $\delta$ slightly below $\delta_c$ 
one may look for a standard scaling form: 
$P_{inv}(t)\simeq t^{-1}g(t/\tau(\delta))$, with $g(x)$ a scaling function 
rapidly decaying to zero at large $x$ and $\tau(\delta)$ 
a diverging timescale as $\delta\rightarrow \delta_c$. The inset of
Fig. \ref{figScalPinv} displays $tP_{inv}(t)$ vs. $t$ for different values 
of $\delta$ near 4.1 and shows that the above ansatz does not hold.

Interestingly, the probability exhibits an unusual behavior instead.
First, the different curves can not be rescaled onto a single curve.
Second, a pure power-law behavior 
was never obtained for $P_{inv}(t)$ (nor $P_0(t)$) for the values of 
$\delta$ considered in this study. Intricate corrections to scaling in 
the form of logarithmic oscillations are observed. 
Log-oscillations have been
observed in a variety of systems and are a manifestation
of the phenomenon of discrete scale invariance \cite{sornette}.
The log-oscillations have a large period, of order $2\ln 10$, 
which complicates the observation of several periods:
we can not conclude whether they converge toward a finite
amplitude or are amplified. However, the amplitude of the 
oscillations is minimum at $\delta_c\simeq 4.1$.  

The leading $1/t$ decay of $P_{inv}$ at $\delta_c$ is probably not a 
coincidence. From a renormalization 
group (RG) perspective, the model has a trivial attracting 
fixed point, $\delta^*=0$, corresponding to simple ballistic motion 
in very narrow strips. The $1/t$ law indicates that 
$\delta_c$ should lie at the boundary of the basin of attraction of that 
fixed point. Besides, the increasing corrections from power-law 
behavior for $\delta$ 
slightly above $\delta_c$ (see Fig. \ref{figScalPinv} at $\delta=5$) suggest 
that RG trajectories above $\delta_c$ flow towards an other attracting 
fixed point (that could be $\delta^*=\infty$). This argument supports 
the idea that the transition is not a cross-over and that $\delta_c$ 
might be a non trivial repelling fixed point. 

 An other possibility is that the walk may become asymptotically
ballistic without inversions after extremely large times, unreachable 
with standard numerical methods. In this case, the results above 
would describe a very long transient preceding an asymptotic regime
of limited practical relevance.

\begin{figure}
\onefigure[width=4.6cm,angle=-90] {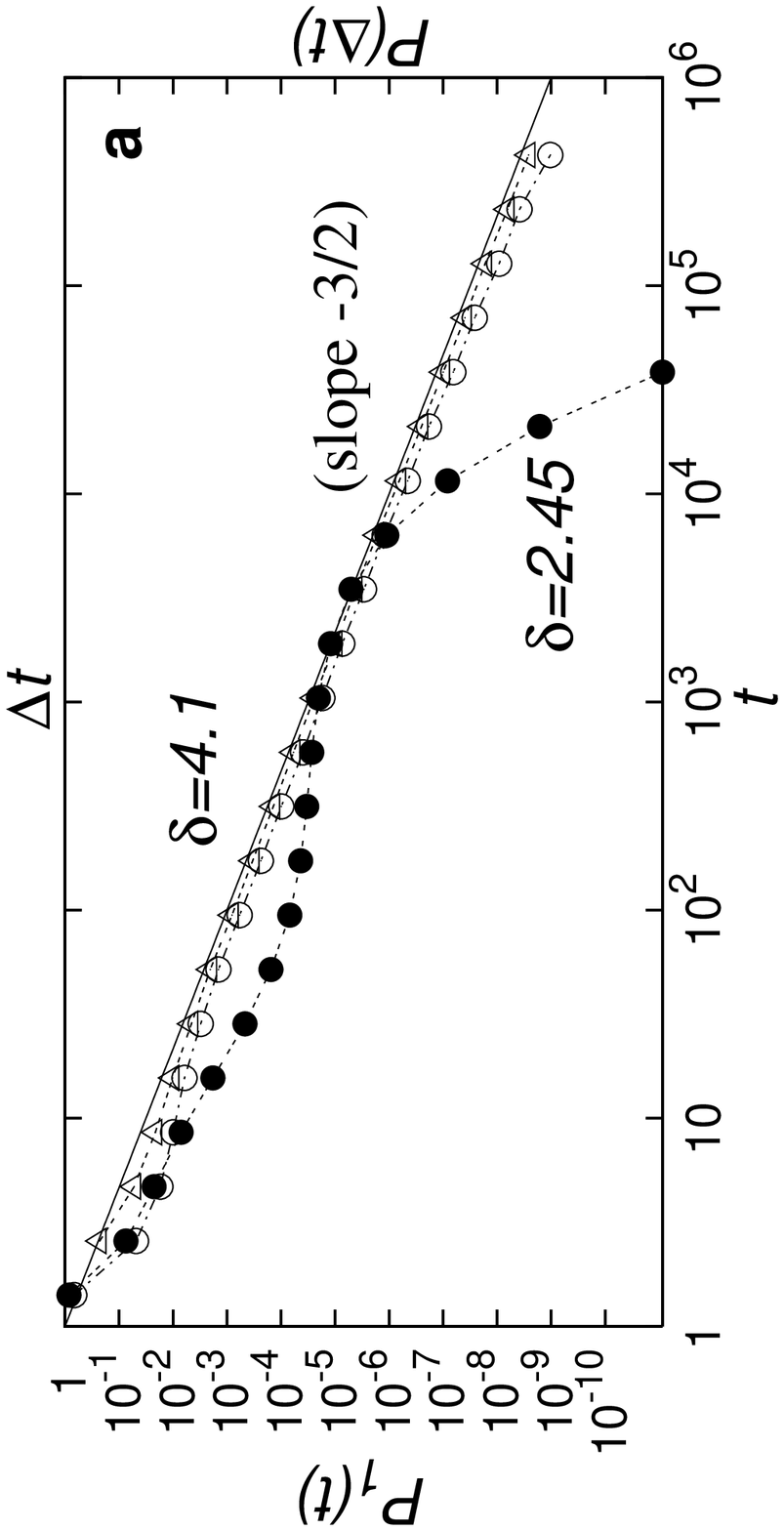}

\vspace{.3cm}
\onefigure[width=4.6cm,angle=-90] {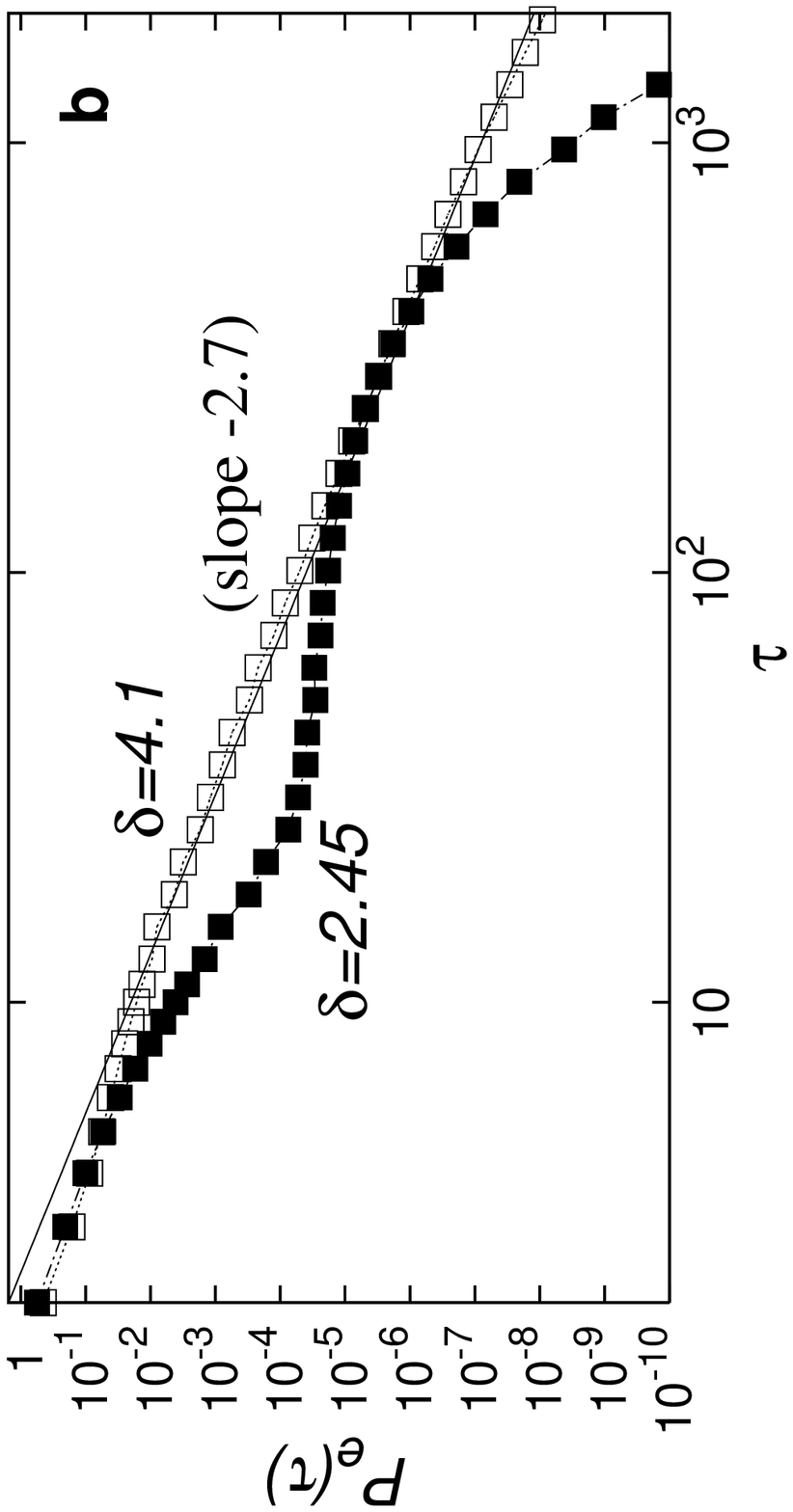}
\caption{ a) Probability distribution $P_1(t)$ of the first-passage time 
at $\delta=4.1$ ($\triangle$). Opposite side: probability distribution 
$P(\Delta t)$ of the time intervals between two consecutive zeros of $N_u(t)$ 
for $\delta=4.1$ ($\circ$) and $\delta=2.45$ ($\bullet$). 
$P_1(t)$ and $P(\Delta t)$ are obtained from $8\ 10^4$ and $10^3$
independent runs respectively. b) Probability distribution
of the excursion duration ($\tau\ge 2)$. Lines are guides to the eye.}
\label{figFirstPass}
\end{figure}

Other insights into inversion processes can be gained from
the distribution $P_1(t)$ of first-passage times.
The first-passage time is defined here as the step number when
the walker crosses for the first time $x_0$.
As shown in figure \ref{figFirstPass}a, a $t^{-\alpha}$ law with 
$\alpha\simeq 3/2$ holds remarkably well over nearly 6 decades
in the vicinity of $\delta_c$.
As for $P_{inv}$ (and $P_0$), log-oscillations 
were detected in $P_1$ at (and near) $\delta_c$.

This exponent value can be qualitatively explained with the help of the simple 
two-state approximation described above, where the probability that the 
velocity changes its sign for the first time at time $t$ reads:
$p_1(t)=p_{inv}(t)\exp[-\int_0^t p_{inv}(t')dt']$. If $p_{inv}\simeq c/t$ 
at large times, then $p_1(t)\sim t^{-\alpha}$ with $\alpha=1+c$. The numerical 
value of $c$ calculated from $P_{inv}$ at $\delta=4.1$ yields 
$\alpha\simeq 1.33$. This value is 
close to, but lower than the observed $3/2$. Therefore,
inversion events are not independent 
but probably long-range correlated. It is actually surprising 
(and most likely coincidental) that the first-passage exponent is close to 
the simple value of the $1d$ random walk \cite{redner}.

\section{Backward excursions and unvisited sites}

We now come back to the description of backward excursions, that are much 
more frequent than inversion events. 
The probability distribution of excursion durations, 
$P_e(\tau)$, can be obtained from the sizes of the time intervals when 
$x_{min}(t)$ or $x_{max}(t)$ remains constant. As shown in 
figure \ref{figFirstPass}b, in the vicinity of $\delta_c$ this distribution 
is also well fitted by a power-law behavior, $P_e(\tau)\sim \tau^{-\beta}$, 
with $\beta\simeq 2.7$. This distribution has finite 
first moment but infinite variance. On average, the walker remains 
\lq\lq trapped" in an excursion during a finite number of steps, but its
progression is quite intermittent.
For strip widths well below $\delta_c$, excursions are still observed and
$P_e(\tau)$ remains fairly broad, although it can no longer be fitted 
with a power-law. 
Generally speaking, backward excursions tend to restore the right-left 
symmetry of the system. For this reason they 
are reminiscent of the effect of thermal fluctuations on a broken symmetry 
phase in equilibrium. 

\begin{figure}
\onefigure[width=6.45cm,angle=-90] {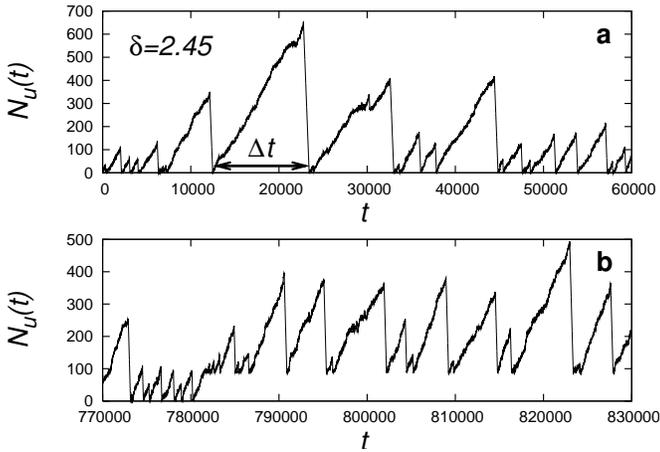}
\caption{Number of unvisited targets $N_u(t)$ as a function of $t$, at early
({\it a}) and late ({\it b}) times, for a same trajectory below the 
transition region ($\delta=2.45)$.}
\label{figNut2.45}
\end{figure}

An other quantity of interest related to excursion dynamics is the number 
of unvisited sites in the explored interval $[x_{min}(t),x_{max}(t)]$, denoted 
as $N_u(t)$. As shown in figure \ref{figNut2.45}a, for values of $\delta$ below 
$\delta_c$, $N_u(t)$ displays cycles of irregular durations analogous to
oscillations in excitable systems. The cycles are composed of $(i)$ 
a slowly increasing part on average and $(ii)$ a fast decay down to zero. 
This behavior reflects the fact that
a small fraction of sites are left as unvisited while the walker 
is drifting in the disordered medium, leading to an increase 
in $N_u(t)$. These sites can be visited later, in a 
long backward excursion, leading to an \lq\lq avalanche-type" relaxation
of $N_u(t)$. (Note that many smaller excursions also occur during the 
ascending part of $N_u(t)$.)
The distribution of the time intervals $\Delta t$
between two successive zeros of $N_u(t)$ is displayed in 
figure \ref{figFirstPass}a. One expects $P(\Delta t)$ and the first-passage 
time distribution $P_1(t)$ to have the same asymptotic behavior, as
observed. At $\delta=4.1$, one finds
$P(\Delta t)\sim (\Delta t)^{-3/2}$, implying that 
$\langle \Delta t\rangle=\infty$: $N_u(t)$ grows asymptotically unbounded.
In the transition region, the walker is therefore unable 
to visit all the targets of the explored interval at large time 
(it is \lq\lq inefficient").
Below $\delta_c$, the distribution $P(\Delta t)$ decays faster 
(figure \ref{figFirstPass}a) and the evolution of $N_u(t)$
seems to have a characteristic cut-off period. In figure \ref{figNut2.45}a, 
where $\delta$ is well below $\delta_c$, this characteristic time is still 
very long  $(\sim 10^4)$. Large avalanches, where
$N_u$ drops from about $600$ to $0$, are present. 

The walker is {\it a priori} \lq\lq efficient" below $\delta_c$, 
since it regularly leaves no sites as unvisited (fig. \ref{figNut2.45}a). 
However, this behavior
is not persistent on very large time-scales, as shown in 
figure \ref{figNut2.45}b. Surprisingly, at a given time that can be
of order $10^6$ or more, $N_u(t)$ does not come back 
to zero and starts to oscillate above a finite value. This happens 
when a large excursion fails to visit some of the unvisited targets 
left behind. After such an incomplete excursion, 
$\langle \Delta t\rangle$ obviously starts to grow with time. 
This behavior is observed in a whole
range of parameter values below $\delta_c$, down to about $\delta\sim 1.30$.
The walk is therefore efficient during a finite
time, until a \lq\lq catastrophic" event with very large 
$\Delta t$ occurs.

One can draw an analogy between these \lq\lq outliers" and
very large earthquakes or financial crisis in other contexts
\cite{sornetteoutliers,sornetteendo}: their 
magnitude can not be explained from a simple extrapolation of the 
distribution $P(\Delta t)$, as displayed in fig. \ref{figFirstPass}a, to 
larger arguments. Here, outliers are unusual
precursory patterns of the power-law distribution (in sharp contrast with
common critical phenomena) and can be attributed to the presence of 
log-oscillations in $P(\Delta t)$:  As noticeable in
figure \ref{figFirstPass}a, after an initial steeper decay, 
the distribution for $\delta=2.45<\delta_c$ very closely approaches the 
critical curve
for a range of values around $\Delta t_1\sim5\ 10^3$, before decaying fast 
again. Although computational limitations do not
allow to observe more log-periods, it is very likely that the next 
oscillation can reach (or even cross) 
the critical distribution, for some values $\Delta t_2\gg\Delta t_1$
(see figure \ref{figScalPinv} for a similar behavior at $\delta=3$).
Therefore, the emergence of very large
intervals of order $\Delta t_2$ between the zeros of $N_u$ becomes 
as probable as at $\delta_c$. This explains qualitatively the time series
of figures \ref{figNut2.45}a-b, composed of many intervals of order $10^3-10^4$
followed by a single one of much larger size, precursor of the
critical regime with diverging $\langle \Delta t\rangle$.

\section{Discussion} 
We have shown that a simple self-avoiding process taking place in a 
confined Poissonian random medium can display complex dynamics and
broad distributions in a wide parameter range. Quenched disorder 
introduces randomness in the model, that otherwise follows simple deterministic 
rules. Similar results as reported here should be observed in a semi-infinite 
strip with the walker initially located at one end, with the difference
that the walker would drift without inversions.

In ref.\cite{redqueen}, the trajectories generated 
by the Red Queen rules can be similar to random walks 
after time scales that depend strongly on the lattice size and 
geometry \cite{redqueen}. In contrast with the 
Red Queen Walk, where sites can be revisited after a very long time, 
our model has infinite memory, leading to intermittent and complex behavior.

The evolution of single species is known to be intermittent and not gradual,
long period of stasis being \lq\lq punctuated" by burst of rapid biological
changes \cite{gould}. Such active periods might be driven by the internal 
dynamics of evolution. According to the fossil record, the number of
genera with a lifetime $\tau$ follow a power-law $N\sim 1/\tau^{\beta}$, 
with $\beta\simeq 2$ \cite{pnas}. The Bak-Sneppen model \cite{bak} considers 
interacting species with high mutation barriers, leading to self-organized 
critical states with $\beta=1.1$ \cite{pnas}. This evolution is slower than
observed because it occurs by collective modes, or avalanches. 
Changes are easier in our model (where explicit interactions are ignored), 
but still intermittent. 
The distribution of time-intervals between successive changes in $x_{max}$,
for instance, is fitted with an exponent $\beta=2.7$. One may
speculate that phenotypic restrictions could play a role 
on the punctuated dynamics of evolution, in addition to species interactions.

Our system does not become critical in an ordinary way.
At a critical width $\delta_c$, the inversion 
probability of the drift velocity decays as a power law with small 
log-periodic corrections. In a first harmonic approximation 
\cite{sornette},
$P_{inv}(t)\simeq ct^{-1}[1+a_1\cos(2\pi \ln t/\ln \lambda)]$,
with $a_1\ll 1$ and $\lambda\approx 100$.
This asymptotic regime is numerically hard to reach, as observed in other 
problems with log-oscillations \cite{gandhi,bernasconi}.
The leading term above precisely represents the law that separates, in analogy 
with a two-state stochastic process, asymptotically ballistic
(non-ergodic) trajectories and walks that keep changing direction 
indefinitely.  

The log-oscillations present in various distribution functions
indicate the presence of a hierarchy of time-scales 
related to each other by a particular
scaling factor $\lambda$, such that $P(\lambda t)\simeq
\lambda^{\alpha}P(t)$.
These oscillations are often displayed by cooperative phenomena 
taking place on hierarchical structures
(spin models near criticality \cite{sornette}, contact 
processes \cite{bab}),
or by random walk models with memory \cite{gandhi,kenkre},
among other examples. 
In ref.\cite{bernasconi}
log-oscillations appear in a simple biased $1d$ random walk model
in a disordered medium containing a small fraction of \lq\lq slow" sites, 
where the walker jumps in the direction opposite to the bias with a 
probability close to one.
These slow sites are somehow analogous to our (dynamically generated)
backward excursions. A crucial ingredient leading to log-oscillations in
\cite{bernasconi} is spatial discreteness, where clusters of slow sites trap
the walker during a time that increases exponentially with the cluster size. 
In our model, no such discreteness is apparent. Instead, inversion events 
are correlated in a complicated way: if the strip is sufficiently narrow, 
an inversion can not occur in the interval $[t_1,2t_1]$ if an 
inversion occurred at time $t_1$.
The time intervals separating inversions might introduce a particular scaling 
factor, although its precise origin is unclear.

Well below the critical region,
the distributions exhibit log-oscillations of irregular amplitudes whose 
maxima can be identified with outliers, that are \lq\lq wild" precursors 
of critical fluctuations. From the above discussion, the characteristic 
size of these events can be roughly extrapolated as being
of order $\lambda^n$ ($n=1,2...$),
and as probable as at criticality.
At $\delta_c$, these specific scales are mixed with all the others 
($a_1\ll 1$), in a practically scale invariant distribution. 
A detailed study the behavior near $\delta_c$ remains to be done. 
Correlation functions ({\it e.g.} velocity) other than $P_{inv}$
might exhibit clearer scaling relations.

In this scenario, it is however clear that outliers 
exist in a wide parameter range and do not even require that the bulk
of the distribution follows a power-law. On the contrary,
they are off-critical events by nature. This property has to be contrasted 
with more common views in seismology, for instance, where outliers 
are either considered as coming from the tail of power-law 
distributions \cite{SOC}, or, in a more refined way, as coming 
from a bump at large sizes in a otherwise power-law distribution of bulk 
events \cite{gilsornette}.

\acknowledgments
Fruitful discussions with H. Larralde, F. Leyvraz, M.G.E da Luz, 
O. Miramontes, A. Robledo and G.M. Viswanathan are gratefully acknowledged.

\end{document}